# High intergrain critical current density in fine-grain (Ba$_{0.6}$K$_{0.4}$)Fe$_2$As$_2$ wires and bulks

J. D. Weiss, C. Tarantini, J. Jiang, F. Kametani, A. A. Polyanskii, D. C. Larbalestier, and E. E. Hellstrom*

Applied Superconductivity Center, National High Magnetic Field Laboratory, Florida State University, Tallahassee, FL, 32310, USA

(*) E-mail: Hellstrom@asc.magnet.fsu.edu
Phone: (850) 645-7489
Fax: (850) 645-7754

The K- and Co-doped BaFe$_2$As$_2$ (Ba-122) superconducting compounds are potentially useful for applications because they have upper critical fields (H$_{c2}$) of well over 50 T, H$_{c2}$ anisotropy γ < 2, and thin film critical current densities $J_c$ exceeding 1 MAcm$^{-2}$ [1–4] at 4.2 K. However, thin-film bicrystals of Co-doped Ba-122 clearly exhibit weak link behavior for [001] tilt misorientations of more than about 5°, suggesting that textured substrates would be needed for applications, as in the cuprates[5,6]. Here we present a contrary and very much more positive result in which untextured polycrystalline (Ba$_{0.6}$K$_{0.4}$)Fe$_2$As$_2$ bulks and round wires with high grain boundary density have transport critical current densities well over 0.1 MAcm$^{-2}$ (SF, 4.2 K), more than 10 times higher than that of any other ferropnictide wire. The enhanced grain connectivity is ascribed to their much improved phase purity and to the enhanced vortex stiffness of this low-anisotropy compound (γ ~ 1-2) compared to YBa$_2$Cu$_3$O$_{7-x}$ (γ ~ 5).

The tendency of grain boundaries of the high temperature superconducting cuprates like YBa$_2$Cu$_3$O$_{7-x}$ to be weak linked has been the major impediment to producing wires needed for their application[7,8]. In the search for new classes of superconductors that might displace Nb-Ti and Nb$_3$Sn, the new Fe-based superconductors also demand attention, even though they seem to share similar non-superconducting parent phases to each other and have low carrier densities[9]. Bulk ferropnictide materials made thus far exhibit low global $J_c$ ($J_c^{global}$) values, some of which can be ascribed to extrinsic factors such as their less than full density, the prevalence of grain-boundary-wetting phases, and cracking[10–15]. As a result, $J_c^{global}$ of randomly oriented ferropnictide wires are typically well below 0.01 MAcm$^{-2}$ [10,16,17], although a recent textured K-doped SrFe$_2$As$_2$ (Sr-122)[18] tape claims about 0.025 MAcm$^{-2}$ at self-field (SF). These values are one to two orders of magnitude less than the local intragrain critical current density $J_c^{local}$ measured in single crystals[19–21]. A potentially serious intrinsic problem is the weak link behavior of grain boundaries (GBs) in Ba(Fe$_{1-x}$Co$_x$)$_2$As$_2$ similar to that observed in the cuprates YBa$_2$Cu$_3$O$_{7-x}$ (YBCO) and Bi$_2$Sr$_2$CaCu$_2$O$_8$ (Bi-2212). However the intergrain critical current density across [001] tilt misoriented grain boundaries ($J_c^{gb}$) decreases less rapidly with increasing grain misorientation angle than in YBCO[5,6,9]. The generality of this result is still unclear though because thin film bicrystals of ferropnictides have so far only been grown for one structure (Ba-122) in only the Co-doped variant due to the difficulty of maintaining proper composition control during deposition. Therefore, the study of polycrystalline samples is still of great importance. Here we report a surprisingly positive result with $(Ba_{0.6}K_{0.4})Fe_2As_2$ bulks and wires made by careful low temperature synthesis showing much higher $J_c^{global}$ than in Co-doped Ba-122 bulks. The very fine grain size (~200 nm), comparable to or smaller than the penetration depth, and the low $H_{c2}$ anisotropy provide a basis for high vortex stiffness. The enhanced phase purity and low anisotropy appear to enable transport critical current densities that are high enough to be interesting for applications.

K-doped Ba-122 polycrystals were fabricated by a low temperature reaction pathway that prevents the formation of secondary phases, like Fe-As, that tend to wet the grain boundaries when higher temperatures are used. An additional benefit of low temperature reactions is that the grain size of the Ba-122 phase is very fine, ~200 nm. Moreover, the bulk forms were easily powdered and made into a wire by the powder-in-tube technique, lengths of which were reacted and characterized. Details of our reaction process are in the methods section.

Careful studies of electromagnetic granularity and the broader superconducting properties were made by SQUID magnetometry for $T_c$, vibrating sample magnetometry (VSM) to measure the global magnetic moment in high fields, magneto-optical imaging to measure local granularity on scales of a few μm, as well as transport critical current measurements to determine the global transport current densities. Microstructures were examined at multiple length scales in a Zeiss 1540 EsB/XB scanning electron microscope (SEM) and a JEOL JEM2011 transmission electron microscope (TEM) to assess the size and distribution of secondary phases, the prevalence of cracking and the Ba-122 phase grain size.

Figure 1 shows the magnetic $T_c$ transition of both wire and bulk in a magnetic field of 2 mT. Both samples show a strong diamagnetic signal with little temperature dependence corresponding to superconducting volume fraction > 90%, indicating strong global screening currents crossing many high-angle grain boundaries. To characterize $H_{c2}(T)$, the bulk resistance was measured by a 4-point method in magnetic fields up to 35 T. The resistivity measurements can be found in figure S1 of the supplementary information. As seen in figure 2a, $H_{c2}(0)$ is estimated above 90 T for K-doped Ba-122, well beyond the highest values obtained for $Nb_3Sn$[22] wires and $MgB_2$[23] thin films and comparable to the K-doped Ba-122 single crystal[4] plotted for comparison. Figure 2b shows the bulk and single crystal (H//ab) data[4] normalized by their respective $T_c$ (bulk: 37.4 K, crystal: 38.7 K). The excellent agreement with single crystal data also confirms that this bulk polycrystal is of high quality since $H_{c2}$ of this compound has been shown to be very sensitive to doping.[4]

Figure 3 shows a transmission electron microscopy (TEM) image of the polycrystalline K-doped bulk. The diffraction contrast in figure 3a clearly shows that the average grain size is approximately 200 nm. The electron diffraction pattern from the selected area of Figure 3a indicates a randomly oriented polycrystalline structure containing many high-angle grain boundaries. The high resolution TEM observation confirms clean and well-connected grain boundaries in a randomly oriented polycrystalline bulk, as shown by a typical grain boundary in figure 3b. However, TEM images do reveal some porosity and secondary phases that obstruct some current flow, indicating room for further process optimization. Supplementary figure S2 shows TEM and the electron diffraction pattern of similar Co-doped Ba-122 material, suggesting that the microstructure is comparable with K-doped 122. Magneto optical (MO) imaging was used to image the local field profile $B_x$ produced by magnetization currents induced by magnetic fields of up to 120 mT applied perpendicular to the bulk sample's surface. The MO images in figure 4 show a roof top pattern of magnetic flux density produced by bulk current flow over the entire ~3 mm-long sample, a length scale that is orders of magnitude larger than the ~200 nm grain size seen in figure 3a. Figure 4a shows only a partial flux penetration due to strong induced currents caused by applying a magnetic field of 120 mT after zero-field-cooling (ZFC) the sample to 6 K. Figures 4b and 4c show uniform fully trapped magnetic flux from applying a magnetic field of 120 mT and then field-cooling (FC) the sample from above $T_c$ to 6 K and 32 K, respectively. The calculated current stream lines for the 32 K FC MO image in figure 4c are shown in figure 4d. Very uniform bulk current flow is still present above 30 K, a testament to the material's electromagnetic homogeneity and large superconducting volume fraction even close to $T_c$. MO images of a cross section of our K-doped wire also show good electromagnetic homogeneity comparable to the bulk material and can be found in supplementary figure S3. In contrast, MO images of other ferropnictide bulks show primarily granular currents indicating little or no bulk current flow[1,14]. Clearly, MO indicates there is significant and well distributed $J_c^{global}$ in our material.

Figure 5 shows $J_c^{magnetization}$ and $J_c^{transport}$ of the wire plotted as a function of applied magnetic fields along with other untextured Fe-based superconducting wires[12,24]. Details about the Co-doped wire can be found in the supplementary information. $J_c^{transport}$, measured for H perpendicular to the wire's length, was calculated from the critical current ($I_c$) using the area of the superconducting cross section of the wire (see figure 5 inset). $I_c$ was determined using the electric field criterion $E_c = 1$ µVcm$^{-1}$ (See supplemental figure S4 for I-V curves). $J_c^{magnetization}$ was calculated from VSM measurements using the Bean model. The good agreement between the transport and magnetization measurements indicates a low contribution of the intragrain $J_c^{local}$ component to the magnetization. At self-field, a high $J_c^{global}$ of over 0.12 MAcm$^{-2}$ is obtained, the highest reported $J_c^{global}$ of any ferropnictide bulk or wire so far. $J_c^{global}$ shows a weak field dependence and maintains a reasonably high value of 0.01 MAcm$^{-2}$ at 12 T. Not only are these values high for ferropnictide materials, but they approach the $J_c^{global}$ values desired for applications.

An SEM image of the cross section of the K-doped Ba-122 wire is displayed in the inset to figure 5. The round wires made by PIT processing are advantageous since they can be made cheaply and applied to traditional designs that depend on electromagnetically isotropic round wires. Since the superconductor is made *ex situ* and the final temperature heat treatment is short and does not exceed 600 °C, sheath materials other than Ag may also be used without significantly degrading the conductor due to chemical reactions that occur with the sheath material at high temperatures. This may allow for the use of stronger, less expensive sheath materials. We have yet to explore in detail the effect of texturing, adding additional dopants, multi-core wiring, or over-doping potassium, all of which have been shown to increase $J_c$ in ferropnictide wires or tapes[11,18,25,26].

We propose that the unexpectedly high $J_c^{global}$ arises through a combination of factors. First, our heat treatment occurs at a sufficiently low temperature that secondary phases do not wet grain boundaries, as earlier studies showed that such phases block current[15]. Second, high pressure synthesis results in nearly 100% dense material, which further contributes to good connectivity. Third, the fine grain size makes planar GBs very rare and the low $\gamma$ value makes the vortex stiffness high. Thus, although essentially all vortices cross GBs, which may have a depressed superconducting order parameter, the GB vortex portion is short and can be anchored by the strong pinning of the superconducting segments lying in the grains. This situation has been studied for YBCO bicrystals with planar GBs, varying the angle between the B vector and the GB plane[9,27]. Only when the two are close and a significant length of vortex lies in the GB is the GB $J_c$ depressed below the intragrain $J_c$ value. In our K-doped Ba-122 bulks and wires, which have very small grains and thus a high density of non-planar GBs plus a small $\gamma$, we may expect that very little of the vortices actually lies in any GB. A final possibility is that, the K-doped compound may have less depressed GB order parameters, perhaps due to a higher carrier density induced by K segregation to the grain boundaries. The higher $J_c^{global}$ of the K-doped wire compared to the Co-doped wire plotted in figure 5 suggests that particular compound-related factors may also be playing a role, whether due to differences in GB properties[28] or to the role of hole (K), rather than electron (Co) doping in Ba-122. Bicrystal experiments on the K-doped Ba-122 will be valuable to explore the specific properties of planar grain boundaries that are possibly less weak linked than in other superconductors.

Methods

The elements were first mixed to obtain nominal composition $(Ba_{0.6}K_{0.4})Fe_2As_2$ or $Ba(Fe_{0.92}Co_{0.08})_2As_2$ and then ball milled for 1 hour. During the milling, an exothermic reaction occurred, partially reacting the material to the Ba-122 phase. The ball milled material was wrapped with Nb foil and placed in a stainless steel ampoule which was evacuated, welded shut, compressed into a pellet with a cold isostatic press (CIP) at 275 MPa, and then heat treated in a hot isostatic press (HIP) under 192 MPa of pressure at 600 °C for 20 hours. The material was re-milled and heat treated again as above for 10 hours to obtain a more homogenous bulk. For the wires, the bulk material was milled after the previous two steps and packed into a Ag tube (6.35 mm OD, 4.35 mm ID). The tube ends were plugged, swaged and welded shut. The tube was then groove rolled, followed by drawing to a 0.8 mm OD wire. Pieces of the Ag-clad wire were sealed in Cu tubing (1.57 mm OD, 0.86 mm ID) under vacuum by welding the ends shut. The Cu tubing was then groove rolled to an OD of ~1.35 mm. The Cu/Ag clad wires were then compressed in a CIP under 2 GPa pressure and heat treated for 10 hours at 600 °C in the HIP, as above.

Acknowledgments

Bill Starch, Muriel Hannion, and Michael Santos provided technical support. This work is supported by NSF DMR-1006584, by the National High Magnetic Field Laboratory which is supported by the National Science Foundation under NSF/DMR-0084173 and by the State of Florida.

Author contributions

Additional information

The authors declare no competing financial interests. Supplementary information accompanies this paper on www.nature.com/naturematerials. Reprints and permissions information is available online at http://npg.nature.com/reprintsandpermissions. Correspondence and requests for materials should be addressed to E.E.H.

Figure captions

**Figure 1 – Volumetric magnetic susceptibility as a function of temperature for K-doped Ba-122 wire and bulk**. The magnetic response was evaluated by warming above $T_c$ after zero field cooling to 5 K and applying a field of 2 mT parallel to the sample's length.

**Figure 2 – Upper critical field as a function of temperature. (a)** $H_{c2}(T)$ defined at 90% resistance for the K-doped Ba-122 bulk compared to an optimally doped single crystal from reference 4, a $Nb_3Sn$ wire from reference 22, and a textured $MgB_2$ thin film from reference 23 with H applied parallel (closed symbols) and orthogonal (open symbols) to its surface. The dotted line is a rescaled fit from reference 4 to guide the eye. **(b)** $H_{c2}$ and temperature normalized by $T_c$ to show close agreement between bulk polycrystal and single crystal with H//ab.

**Figure 3 – Microstructures of K-doped Ba-122 bulk investigated by TEM. (a)** TEM image of polycrystalline bulk K-doped Ba-122 material showing several equiaxed grains with average grain diameter of ~200 nm. Inset is a selected area electron diffraction image of **a** that indicates that the grains of the material are randomly oriented with many high-angle grain boundaries.

**(b)** HRTEM image of a typical K-doped Ba-122 grain boundary where the TEM sample was tilted so the electron beam was almost parallel to the GB plane. The lattice fringes of upper and bottom grains meet at the grain boundary without an amorphous contrast, indicating the grain boundary is clean without a wetting impurity phase.

**Figure 4 – Magneto-optical images of a rectangular piece of K-doped Ba-122 bulk material with magnetic fields applied perpendicular to plain of the sample (thickness = 0.7 mm). (a)** Magneto-optical image of partial flux penetration after zero-field-cooling (ZFC) the sample to 6 K and applying a magnetic field of 120 mT. **(b)** Magneto-optical image of trapped magnetic flux in a sample field-cooled (FC) to 6 K in an external magnetic field of 120 mT. **(c)** Magneto-optical image of trapped magnetic flux in a sample FC to 32 K in an external magnetic field of 120 mT. **(d)** Current stream lines calculated for c that illustrate the uniform current distribution that circulates inside the bulk even near $T_c$.

**Figure 5 -** $J_c^{transport}$ **(symbols) and** $J_c^{magnetization}$ **(solid lines) as a function of applied magnetic field at 4.2 K for the K-doped wire compared to other round, untextured, Fe-based superconducting wires**. Sm-1111 wire is from reference 13 and FeSe wire is from reference 24. Inset is an SEM image of the K-doped mono-core wire showing the round cross section with Ag and Cu sheaths.

J. D. Weiss et al.J. D. Weiss et al.

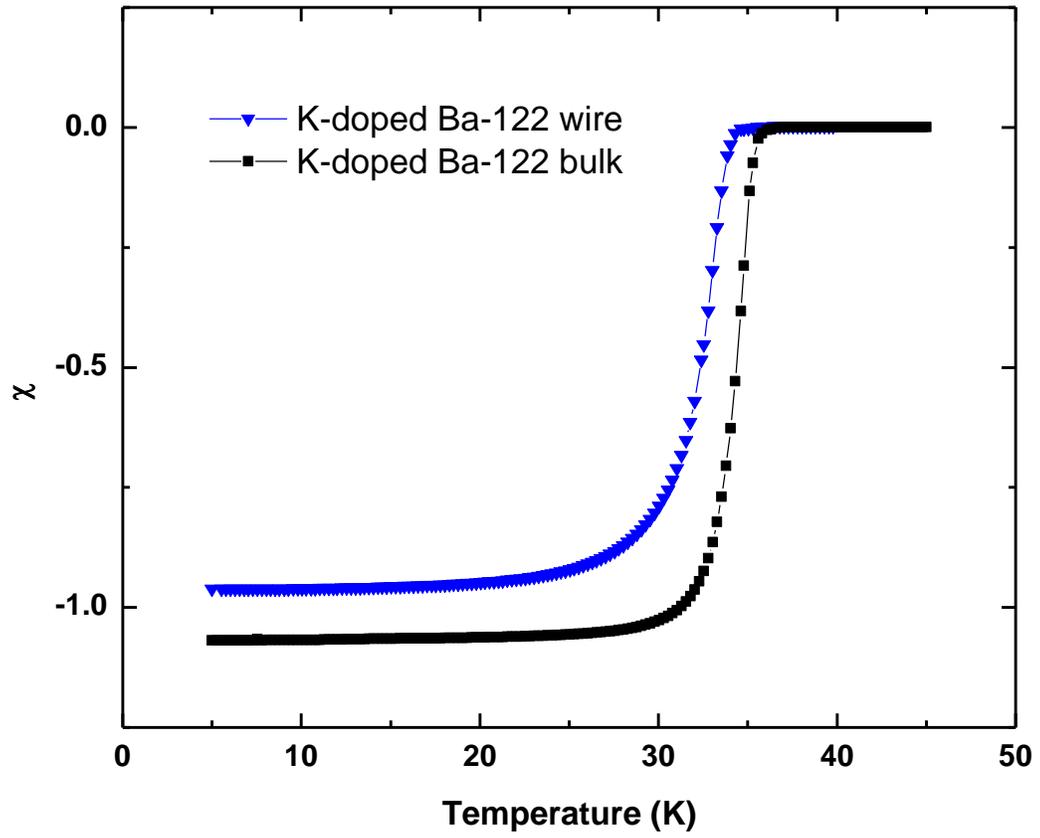

Figure 1

1111

J. D. Weiss et al.

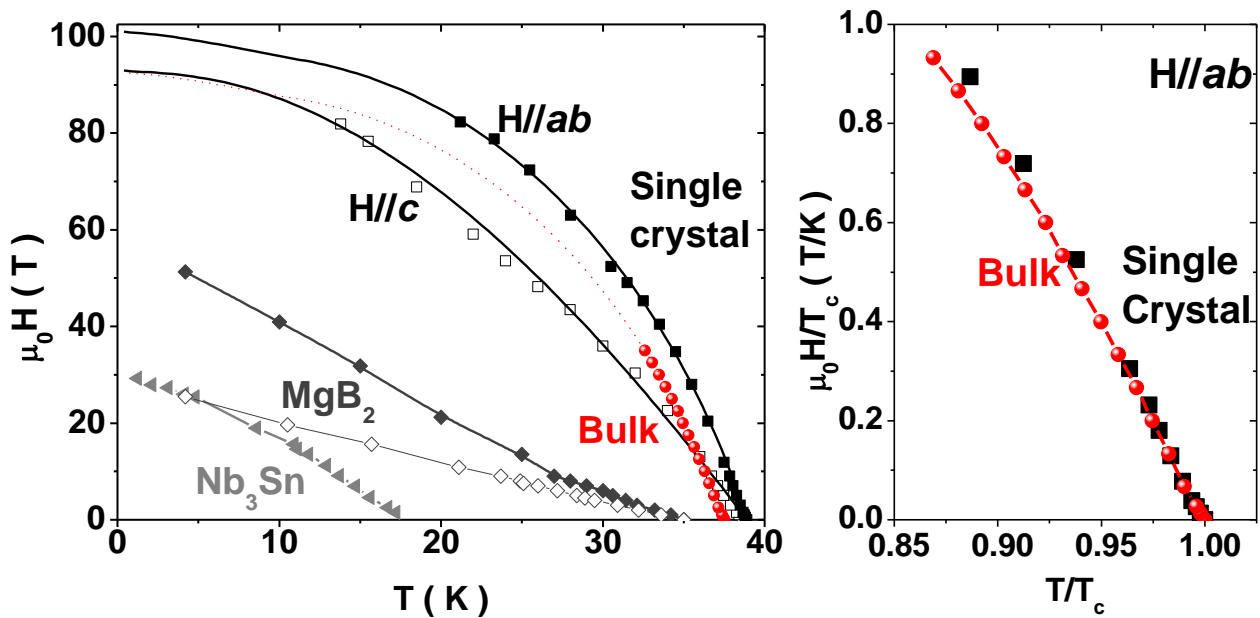

Figure 2

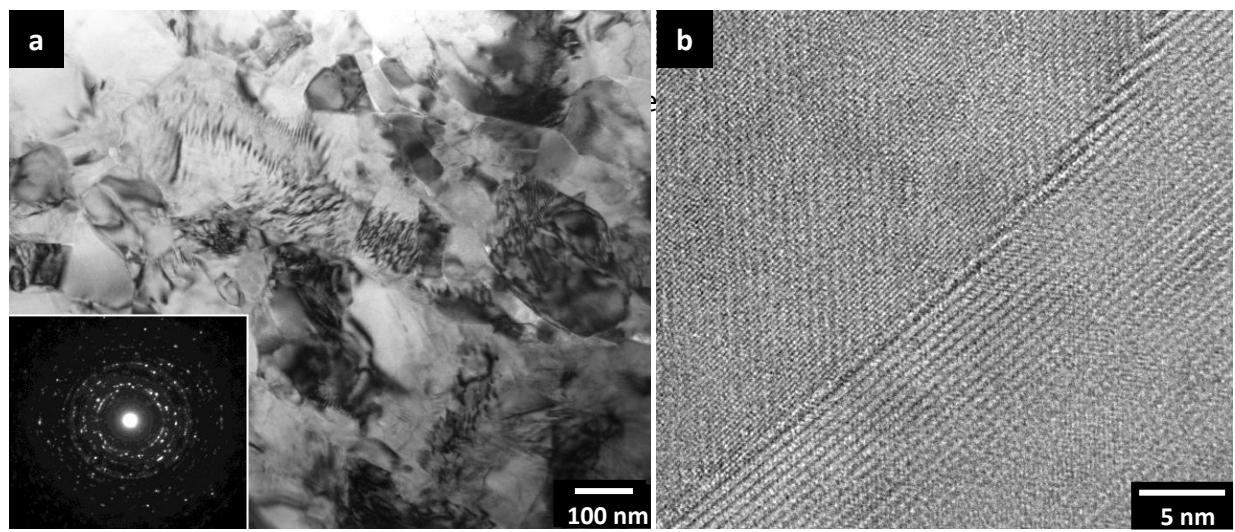

Figure 3

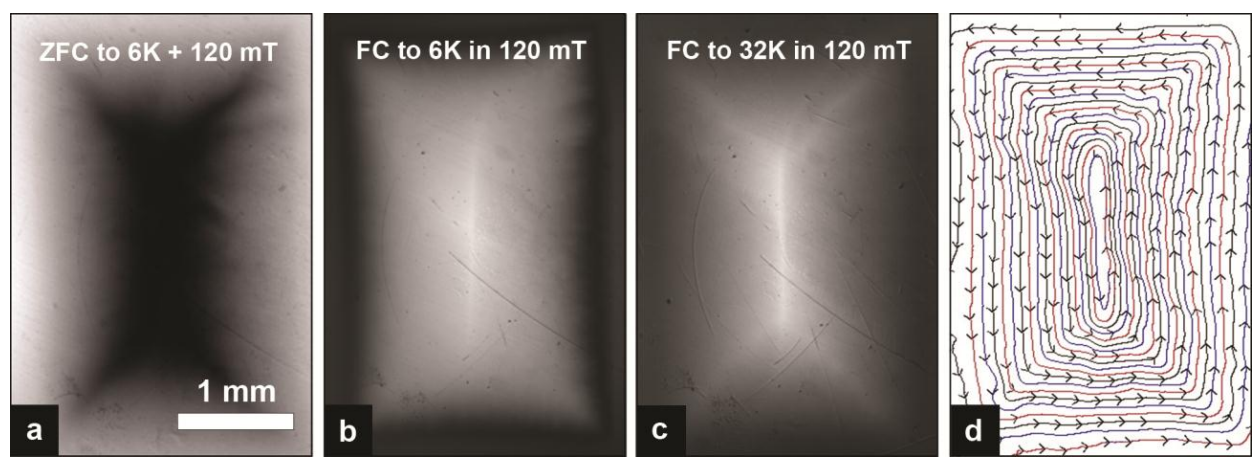

Figure 4

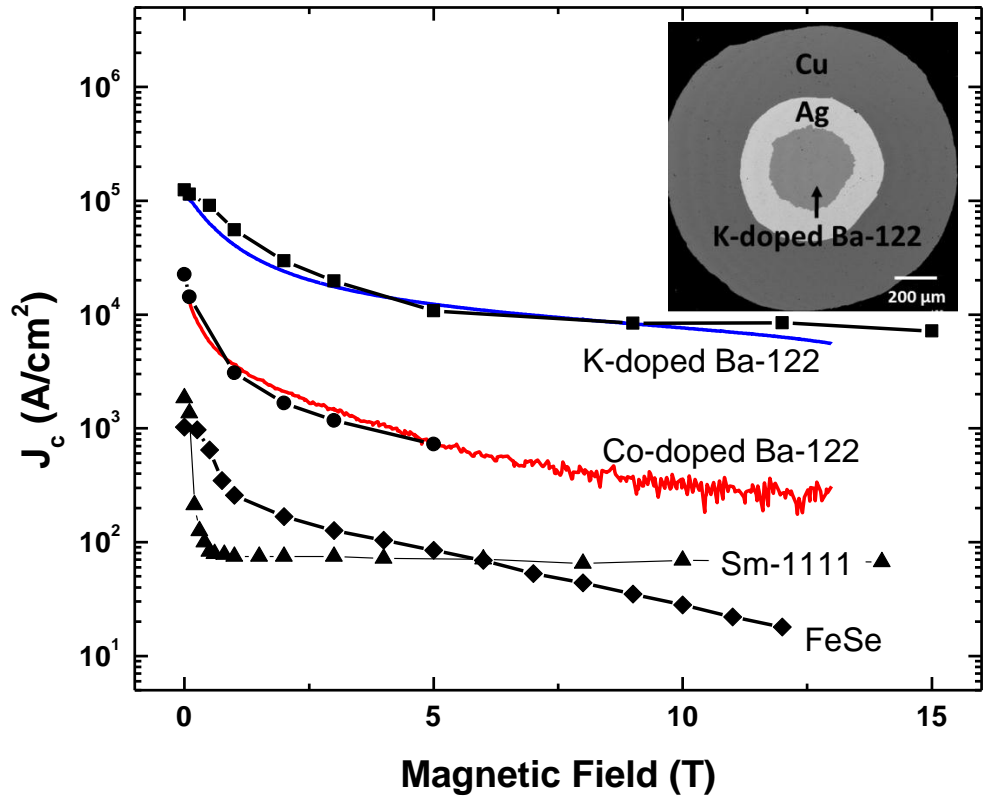

Figure 5

**Supplementary Information**

**High intergrain critical current density in fine grain (Ba$_{0.6}$K$_{0.4}$)Fe$_2$As$_2$ wires and bulks**

J. D. Weiss, C. Tarantini, J. Jiang, F. Kametani, A. A. Polyanskii, D. C. Larbalestier, and E. E. Hellstrom*

Applied Superconductivity Center, National High Magnetic Field Laboratory, Florida State University, Tallahassee, FL, 32310, USA

(*) E-mail: Hellstrom@asc.magnet.fsu.edu
Phone: (850) 645-7489
Fax: (850) 645-7754

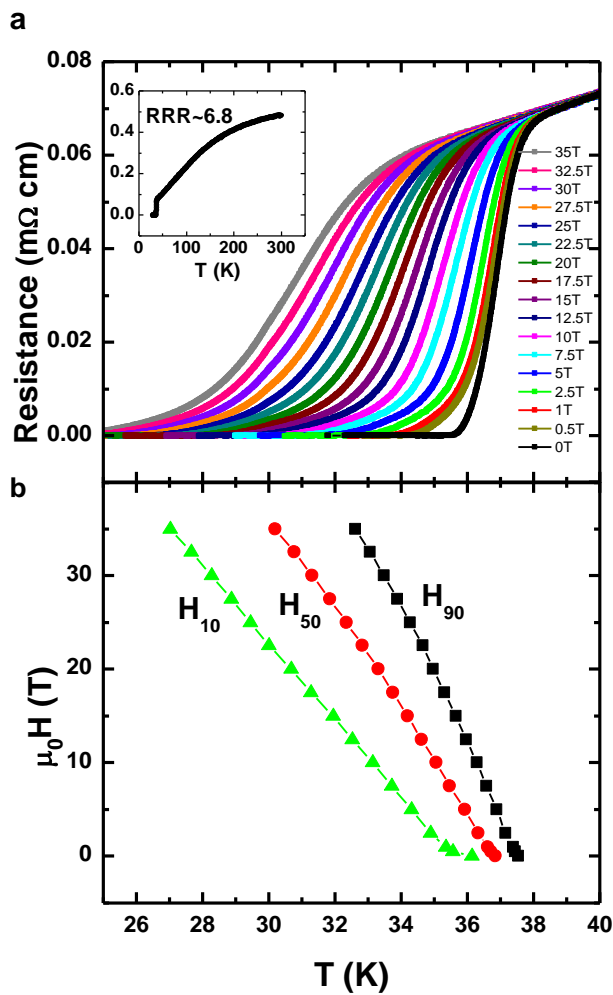

**Figure S1 – Resistivity measurements of K-doped Ba-122 bulk material. (a)** Temperature dependence of resistance at different magnetic fields up to 35 T. The trend of resistivity with respect to applied field is very similar to that of K-doped single crystals , even though our bulk is untextured.  It has ρ(300 K & 39 K) = (0.48 mΩcm & 0.07 mΩcm) compared to ρ(300 K & 39 K) = (0.6 mΩcm & 0.05-0.12 mΩcm) for single crystals REF, indicating that the normal-state properties are not being degraded by the presence of grain boundaries.  Inset is 0 T resistivity up to 300K and RRR is ρ(300 K) divided by ρ(39 K). **(b)** $H_{c2}(T)$ defined at 90% ($H_{90}$), 50% ($H_{50}$) and 10% ($H_{10}$) resistance.

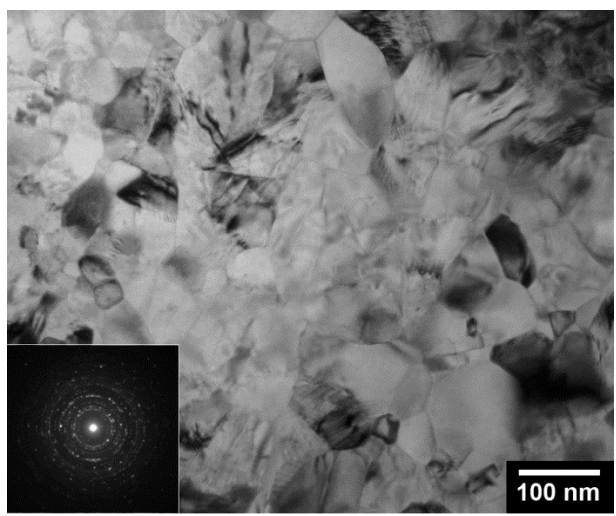

**Figure S2 – Microstructure of Co-doped Ba-122 wire investigated by TEM.** TEM image of polycrystalline bulk Co-doped Ba-122 material showing equiaxed grains with average grain diameter less than 200 nm. Inset is a selected area electron diffraction image that indicates the grains of the material are randomly oriented with many high-angle grain boundaries. TEM confirms the Co-doped wire is structurally comparable to the K-doped wire with many well connected grains.

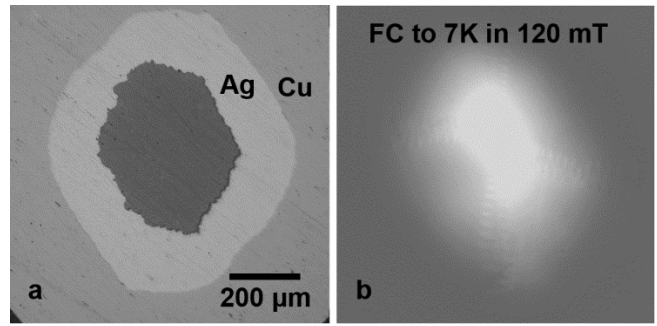

**Figure S3 – Optical and Magneto-optical images of a K-doped Ba-122 wire cross section with magnetic fields applied perpendicular to the shown cross section. (a)** Optical image of the wire cross section showing superconducting core surrounded by Ag and Cu sheath. **(b)** Magneto-optical image of trapped magnetic flux in the wire field-cooled (FC) to 7 K in an external magnetic field of 120 mT.

J. D. Weiss et al.

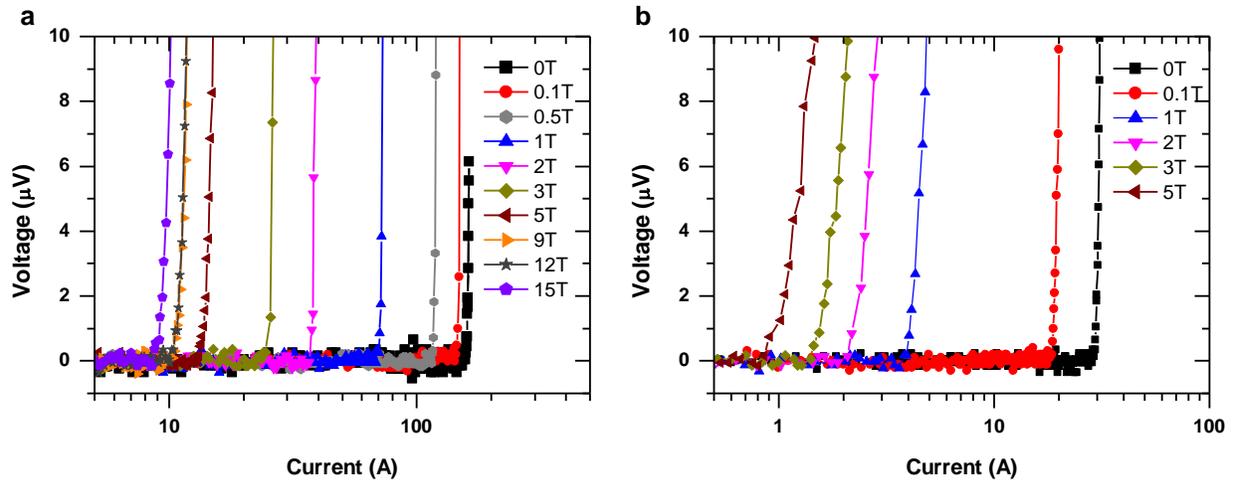

**Figure S4 – I-V curves for the Ba-122 wires at different fields. (a)** K-doped Ba-122 wire measured in fields up 15 T. **(b)** Co-doped Ba-122 wire measured in fields up to 5 T. The Co-doped wire was made by the same PIT process used for the K-doped wire. Voltage response was measured at 4.2 K on a 4 cm pieces of wire with voltage taps approximately 1 cm apart. Measurements were made with increasing current.